\documentclass[doublecol]{epl2} 

\title{The extensive nature of group quality}
\shorttitle{The extensive nature of group quality} 

\author{R. Kenna\inst{1} \and B. Berche\inst{2}}

\institute{                    
  \inst{1} Applied Mathematics Research Centre, Coventry University, Coventry, CV1 5FB, England, EU\\
  \inst{2} Institut Jean Lamour\footnote{Laboratoire associ\'e au CNRS UMR 7198},
 CNRS -- Nancy Universit\'{e} - UPVM, B.P. 70239, F - 54506 Vand{\oe}uvre l\`es Nancy Cedex, France, EU
}

\abstract{
We consider groups of interacting nodes engaged in an activity 
as many-body, complex systems and analyse their cooperative behaviour 
from a mean-field point of view. 
We show that
inter-nodal interactions rather than accumulated individual node strengths 
dominate the quality of group activity, 
 and give rise to 
phenomena akin to phase transitions, where the extensive relationship between group
quality and quantity reduces. 
The theory is tested using empirical data on quantity and quality of scientific research groups, for which
critical masses are determined.
}

\begin{document}

\maketitle

\section{Introduction}

In recent times  statistical physics  has found applications beyond its 
traditional confines and its methods have been deployed to garner new insights in many academic disciplines
\cite{Galam}. 
These include sociology, economics, complex networks  \cite{networks} as well as in more exotic areas \cite{football}.
Each of these disciplines involve cooperative phenomena emerging from the interactions between individual units.
Microscopic physical models  --~mostly of a rather simple nature~-- help explain how the  
properties of such complex systems arise from the  properties of their individual parts. 

Here we consider groups of interacting nodes engaged in a common activity, 
such as research groups of interacting scientists, as complex systems. 
Also in recent times, the assessment of the relative strengths of such research groups has grown in importance
as universities, funding councils and governments seek to decide on where to focus investment.
The debate amongst policy makers is whether to concentrate funding in relatively few well resourced
institutions or to promote competition amongst a wider set, where pockets of research excellence are found \cite{Ha09}.
A central question in this debate is whether there exists a critical mass in research, and if so, what is it?

We present a simple model which captures the complex nature of group quality 
as deriving both from the strengths of its members {\emph{and}} from the interactions between them.
We show the latter is, in fact, the dominant mechanism which drives the quality of group activity.
Phenomena akin to phase transitions are manifest in our model and 
two related significant group sizes emerge: 
a critical mass below which a group is vulnerable to extinction 
and a higher value at which  the correlation between group quality and quantity reduces.
We test our theory using empirical data on the quality of research groups
in natural sciences, for which we determine the critical masses.

\section{Group quality and quantity}

The empirical work presented here is based upon measures of research quality as determined by the UK's
Research Assessment Exercise (RAE) and the French equivalent, which is performed by 
{\emph{Agence d'{\'E}valuation de la Recherche et de l'Enseignement Sup{\'e}rieur}\/} (AERES). 
For the RAE, research groups were scrutinized to  determine the proportion of work  
which fell into five quality levels from  4* (world leading) to 1* (nationally recognized) and unclassified. 
Based on the resulting quality profiles, a formula 
is used to determine how research funding is distributed. 
Assuming this to be a reasonably reliable and robust process, we
consider this as a basis for a 
measurement of the quality  of research groups. 
If $p_{n*}$ represents the percentage of  a team's 
research which was evaluated as $n$*, this gives that team's  overall quality as
$
 s = p_{4^*} + 3 p_{3^*}/7 + p_{2^*}/7
$.
A naive expectation is that, while larger quality values may be associated with older,
more prestigious institutions, these variables are otherwise randomly distributed. 
Indeed, this is the picture implied by Fig.~1(a), where quality measurements
are plotted alphabetically for the 45 UK applied mathematics  (which includes theoretical physics) groups assessed at RAE.
Such plots form the basis of rankings through which the various institutions are compared.
\begin{figure}[t]
\begin{center}
\includegraphics[width=0.65\columnwidth, angle=0]{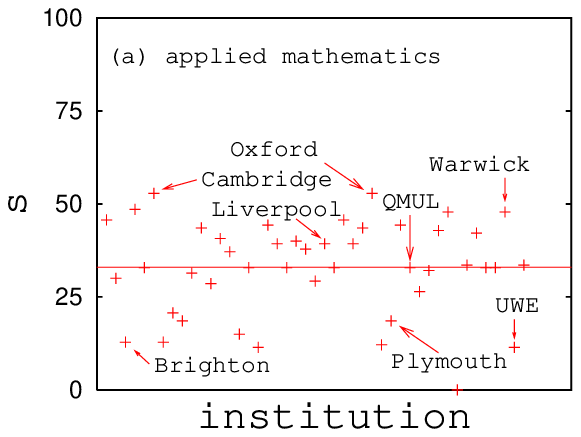}\\
\includegraphics[width=0.65\columnwidth, angle=0]{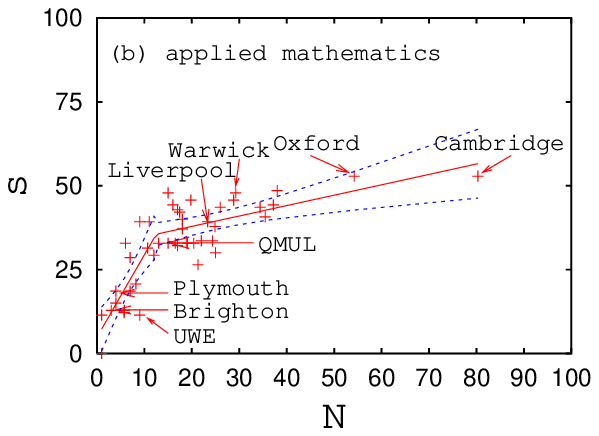}
\caption{Research quality $s$ for each of 45 UK applied mathematics research groups
(a) arranged alphabetically and (b) as a function of group size $N$.
(The sample institutions include Queen Mary, University of London (QMUL) and 
the University of the West of England Bristol (UWE).)
In (b) the solid lines are
piecewise linear regression best-fits to the data and the  
dashed curves represent 95\% condidence intervals for these fits.
}
\end{center}
\label{fig1}
\end{figure}
Associated with these rankings is the notion that group quality is a measure 
of the average calibre of individuals forming the group, i.e., if ${a_g}_i$
represents the strenth of the $i^{\rm{th}}$ individual in group $g$, then
the strength of that group is naively given by $S_g = \sum_{i=1}^{N}{{a_g}_i} = N \bar{a}_g $,
where $ \bar{a}_g$ is the average strength of the $N$ individuals in the group.
Defining the {\emph{quality}} $s_g$ of the group as its average strength per head, so that $s_g = S_g/N$,
one is then led to the naive conclusion that mean calibre $\bar{a}_g$ is given by the group quality $s_g$.
This is the standard conclusion from a naive approach associated with Fig.~1(a).
We will show that such a conclusion is dangerous and wrong.

A hint at the correct interpretation is given in Fig.~1(b), where quality measures are 
plotted against {\emph{quantity}} in the form of group sizes. Clearly there is a linear relationship between 
quality $s$ and quantity $N$, at least in the left part of the curve. 
From a statistical physics point of view, there is an obvious interpretation in terms of the 
links between individuals. This interpretation will lead to strong and unobvious conclusions 
regarding the relationship between quality and quantity as well as to the quantification of the 
hitherto intuitive notion of critical mass in research \cite{Ha09}.

We first consider the situation where the number of nodes $N$ 
in group $g$ is not too large, and where one may represent the group as a graph 
which is complete in the sense that it has $N(N-1)/2$ active edges.
With 
${b_g}_{i,j}$ representing the strength of interaction between the $i^{\rm{th}}$ and $j^{\rm{th}}$
individuals, the group strength is
$
 S_g = \sum_{i=1}^{N}{{a_g}_i} + \sum_{\langle{i,j}\rangle=1}^{N(N-1)/2}{{b_g}_{i,j}} 
= N \bar{a}_g 
+ N(N-1) \bar{b}_g/2 ,
$
where $\langle{i,j}\rangle$ represents the link between nodes $i$ and $j$ and
 $\bar{b}_g$ is the average strength of interactions between them.
Inspired by molecular field theory  \cite{Weiss}, 
we now write the average total strength of such complete single-cluster groups as
\begin{equation}
 S \propto  \bar{a} N + \frac{\bar{b}}{2}N(N-1)
,
\label{smallmedium}
\end{equation}
where $\bar{a}$ and $\bar{b}$ are the mean values of node and interaction 
strength averaged over all groups.

In fact, since two-way communication can only be  carried out effectively
between a limited number of nodes, it may be further expected that, as 
the size $N$ of a group is increased, a  transition point $N_c$ 
is eventually reached, beyond which the clustering coefficient of the graph decreases.
If a given node can interact meaningfully with at most $N_c$ others, the 
group may cluster into $ N/(\alpha N_c)$ {\emph{subgroups}, of mean size $\alpha N_c$, say, 
within which meaningful interactions can take place. 
This is reminiscent of the de Gennes blob picture for polymers \cite{deGennes}.
Mean strength of the collective due to individual strength and intra-subgroup interaction is now
 $S=\bar{a} N +\bar{b}N(\alpha N_c - 1)/2$. 
Each of the $ N/(\alpha N_c)$ subgroups may interact with strength $\beta$, say, with the
 $ N/(\alpha N_c)-1$ others. This inter-subgroup interaction contributes
an additional strength proportional to the number of such inter-subgroup links. 
The mean strength of a group with $N>N_c$ is then
\begin{equation}
 S \propto  \bar{a} N + \frac{\bar{b}}{2}N(\alpha N_c-1) + \frac{\beta}{2}\frac{N}{\alpha N_c}
\left({\frac{N}{\alpha  N_c}-1 }\right)
.
\label{large}
\end{equation}
The size $N_c$ may thus be considered a transition point between 
``small/medium'' and ``large'' groups. It  marks the number of nodes with which an
individual can cooperate in a meaningful sense.

\begin{figure*}[t]
\begin{center}
\includegraphics[width=0.65\columnwidth, angle=0]{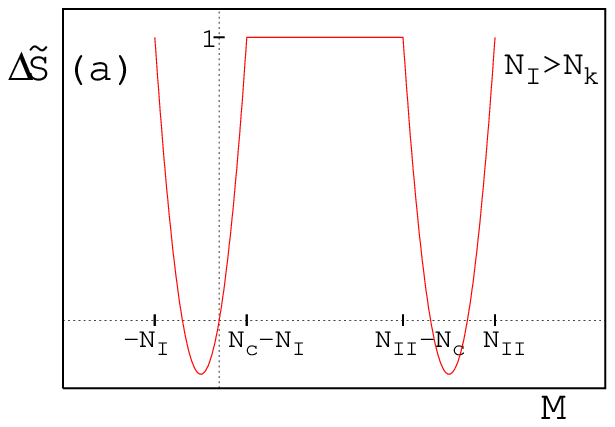}
\includegraphics[width=0.65\columnwidth, angle=0]{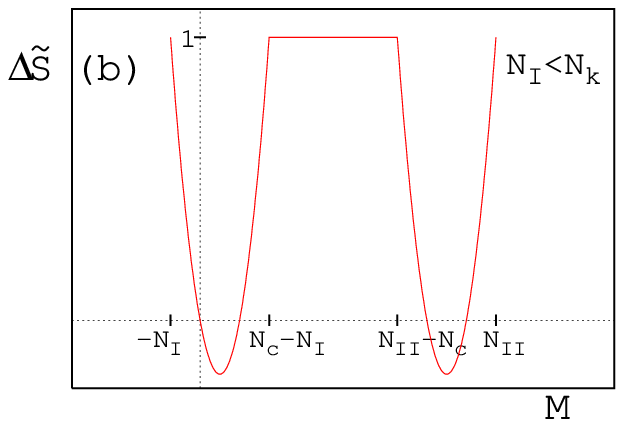}
\includegraphics[width=0.65\columnwidth, angle=0]{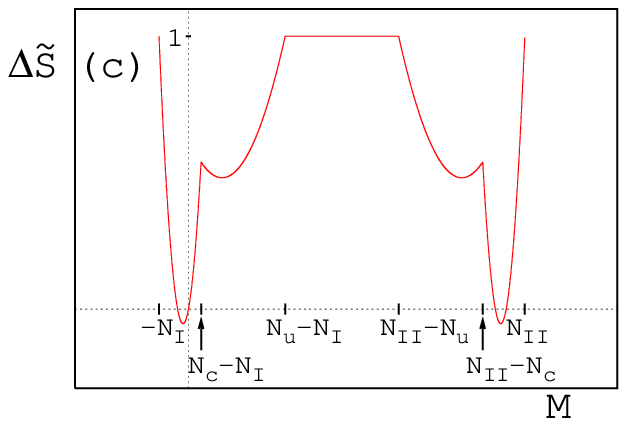}
\caption{Phase diagrams representing the mean total increase in strength in an activity obtained  
by transferring $M$ nodes from group~II to group~I.  
(a) If group~I is  initially supercritical, the positive gradient at $M=0$
indicates that incremental transfer to it  from the large group is globally beneficial for the activity.
(b) If  group~I is initially sub-critical the transfer of mass to it 
from the large group is only beneficial if a sufficient number of nodes 
(namely $N_c-N_{\rm{I}}$) is involved. 
(c) Phase diagram when two breakpoints are present, one at $N=N_c$ and the other at $N=N_u$.
}
\end{center}
\label{canonical}
\end{figure*}

Therefore the ``mean-field''  relationship between the quality of a group engaged in a given activity 
and its quantity or size $N$ is
\begin{equation}
 s = \frac{S}{N} =   \left\{ \begin{array}{ll}
             a_1 + b_1 N &  {\mbox{if $N \le N_c$}} \\
             a_2 + b_2 N &  {\mbox{if $N \ge N_c$}},
             \end{array}
     \right.
\label{Nc}
\end{equation}
where $a_1, \dots, b_2$ are related to the mean individual, intra-group and inter-subgroup 
interaction strengths.  
Continuity at the breakpoint requires that 
\begin{equation}
 N_c = \frac{a_2-a_1}{b_1-b_2}.
\label{Fri}
\end{equation}}
As its size grows beyond $N_c$, two-way collaboration between {\emph{all}} nodes 
is no longer the dominant driver of quality of the group. 
Interactions between subgroups may instead be expected to lead to a milder dependency of $s$ on $N$.
Indeed, from ({\ref{large}), one may expect the slope $b_2$ to decrease  with 
the size of the breakpoint $N_c$. 
We shall provide empirical evidence in support of this and see that $b_2$ is close
to zero for large $N_c$. 
Thus the slope to the right of the breakpoint $N_c$  plays a role similar to that of the 
order parameter in the theory of critical phenomena
(we are not interested in the thermodynamic limit here).

\section{Critical mass}

Having quantified the notion of  ~{\emph{large}} teams as those whose size exceeds the 
transition point $N_c$, we next 
attempt to pin down the meaning of the term {\emph{critical mass}\/}. 
This is loosely described as the value $N_k$ of $N$ beneath which 
groups are not viable in the longer term.
We refer to teams of size $N < N_k$ as {\emph{small}} and groups with  
$N_k \le N <N_c$ as {\emph{medium}} in size.  
To determine $N_k$, we  consider  groups as canonical or 
grand canonical ensembles.

In the latter case, one may ask,  if  new nodes become available, 
is it more  beneficial for the {\emph{strength of the entire society}} in the given activity
 to allocate them to a small, medium or large team?
From (\ref{Nc}), the gradient of $S=sN$ is 
$a_j + 2b_j N$, where $j=1$ or $2$ according to whether $N<N_c$ or $N>N_c$.
The former has greater value, indicating it is more beneficial if these 
new nodes are allocated to the small/medium group, provided $a_1 + 2b_1 N > a_2 + 2b_2 N$ or 
$
 N > N_k
$,
where $N_k$ is given by the scaling relation
\begin{equation}
 N_k =  \frac{a_2 - a_1}{2(b_1-b_2)} = \frac{N_c}{2}
\,,
\label{Nk}
\end{equation}
having used \ref{Fri}. The quantity $N_k$
may then be considered the critical mass for the given activity.

For the canonical approach, we may ask the complementary question: if the total number of nodes
associated with the given activity is fixed, what is the best strategy, 
on average, for transferring them between small/medium and large groups?
We suppose that group~I initially has  $N_{\rm{I}}$ nodes and 
group~II starts with $N_{\rm{II}}$ members. 
Three scenarios arise.
 In scenario~A,  group~I is small/medium  and group~II is large; 
 in scenario~B, both groups are large and Scenario~C is the inverse of Scenario~A and needs no further consideration.
 We consider the transfer of $M$ nodes from group~II to group~I.
 Starting with Scenario~A, the post-transfer strength of the new configuration is given by
$S_i(M) = a_i(N_{\rm{I}}+M ) +b_i (N_{\rm{I}}+M)^2 + a_2(N_{\rm{II}} - M)  + b_2 (N_{\rm{II}}-M)^2,
$
where $i=1$ or $i=2$ for  scenarios~A and~B as the final configuration, respectively.
Measuring the {\emph{total}\/} increase in strength due to the transfer by
\begin{equation}
 \Delta{\tilde{S}}(M) = \frac{S_i(M)  - S_1(0)}{b_1 N_{\rm{I}} (N_c-N_{\rm{I}})},
\label{deltaStilde}
\end{equation}
and plotting this in Fig.~2(a) and (b), up to corrections of order $b_2$, two outcomes emerge.
For scenario~A  the
collective increase $\Delta{\tilde{S}}$ is maximized at  $M = -N_{\rm{I}}$, which represents the
assimilation of group~I into group~II.
For scenario~B,  one finds that this maximum holds
for $N_c-N_{\rm{I}} \le M \le N_{\rm{II}} -N_c$, which represents any configuration in which each group
has at least  $N_c$  members. 

Incremental transfer of staff from group~II to group~I is governed by scenario~A, the outcome of  which
is determined by the gradient of $\Delta{\tilde{S}}$, at the initial point $M=0$,
\begin{equation}
 \left.{\frac{\partial \Delta {\tilde{S}}}{\partial M}}\right|_{M=0} = \frac{2}{N_{\rm{I}}} \frac{N_{\rm{I}} - N_k}{N_c-N_{\rm{I}} }.
\label{slope}
\end{equation} 
Such a move is globally beneficial if this slope is positive. 
This occurs when $N_{\rm{I}} > N_k$, so that group~I is initially supercritical in size.
I.e., it is sensible to promote medium-size groups at the expense of large ones.
If $N_{\rm{I}} < N_k$ (group~I is initially subcritical), the shortest  route to maximal performance is 
to conglomerate all nodes into the large group. Instead, and to avoid extinction, 
group~I should strive to  achieve critical mass, in which case the previous argument prevails.

\begin{figure*}[t]
\begin{center}
\includegraphics[width=0.65\columnwidth, angle=0]{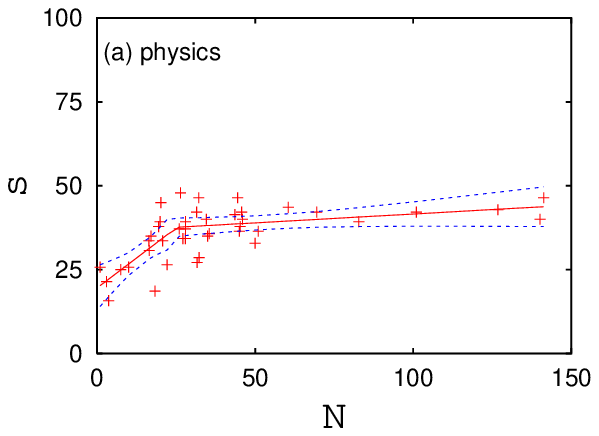}
\includegraphics[width=0.65\columnwidth, angle=0]{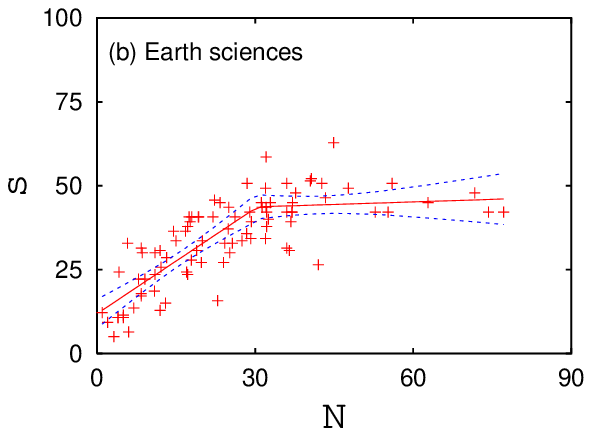}
\includegraphics[width=0.65\columnwidth, angle=0]{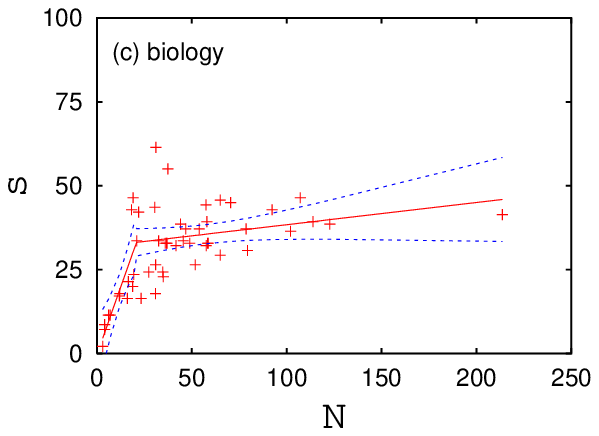}
\caption{Success rate or research quality $s$ as a function of group size $N$
for (a) physics, (b) Earth science and (c) biology.
As in Fig.1(b),  the solid lines are
piecewise linear regression best-fits to the data and the  
dashed curves represent 95\% condidence intervals for these fits.
}
\end{center}
\label{naturalsciences}
\end{figure*}

\section{Application to research groups}

To test our theory, we require a collection of small, medium and large groups
engaged in an activity in which the quality of each group is measured quantitatively.
As mentioned, the UK's academic communities provide suitable testing arenas as the quality of research groups
in many disciplines has been measured through RAE. 

A piecewise linear regression analysis is applied to data sets corresponding to 
measured quality of research groups, fitting to the form (\ref{Nc}). 
In Fig.~3 the data for physics (which includes experimental physics), Earth sciences and biology 
are presented.
For these, as in Fig.~1(b) for applied mathematics,
the positive correlation between  quality and team size reduces beyond a discipline-dependent transition point.
The statistical $P$-values for the null hypothesis that  there is no underlying correlation
were less than $0.1\%$ in each case while the $P$-values for coinciding slopes on 
either side of the transition point are smaller than $0.3\%$.
These small values indicate that we can reject these hypotheses 
for these disciplines, in favour of the alternative that the model 
is sound and that the transition points exist.
The solid curves in the figures are the piecewise linear-regression fits to 
the data and the dashed curves represent the 
resulting 95\% confidence belts. 
For physics, we estimate the critical mass as $N_k = 12.7 \pm 2.4$, about twice the estimate
$N_k = 6.2 \pm 0.9$ for applied 
mathematics  and closer to the estimate $N_k = 15.2 \pm 1.4$ for Earth sciences
and to $N_k = 10.4 \pm 1.6$ for biology.

The measured values for the slopes to the right of the breakpoint are 
$b_2 = 0.05 \pm 0.03$, $0.05 \pm 0.11$ and $0.07 \pm 0.04$ for physics,
Earth sciences and biology, respectively. As expected for large-$N_c$ communities,
these slopes are small compared to that of applied mathematics, which has $b_2= 0.3 \pm 0.1$
and a relatively small value of $N_c$.

In pure mathematics  (see Fig.~4(a)) no transition point was detected and the data is best fitted by a single line,
the intercept ($28.1 \pm 2.8$) and slope ($0.5 \pm 0.2$) 
of which are comparable to the corresponding values for {\emph{large}} 
applied mathematics groups ($a_2=31.7 \pm 12.8$ and $0.31 \pm 0.09$, respectively).
These indicate that these data may also be interpreted as belonging
to {\emph{large}} groups. Then $N_c$ for pure mathematics
may be interpreted as being less than or equal to the size of the
smallest group whose quality was measured, which was $4$, so that $N_k \le 2$. 
This suggests that local cooperation is less significant in pure mathematics, where the work pattern
is more individualized. This is consistent with experience: papers in pure mathematics tend to be authored
by one or two individials, rather than by larger collaborations. 
The results for chemistry are presented in Fig.~4(b), where there is also an
absence of  small groups leading to a relatively large error in the critical mass estimate $N_k=18 \pm 7$.

The above work is based upon the measures of research quality as determined in the UK.
To check its broader generality, we  compare the results of the RAE with those of the French equivalent,
which is performed by the {\emph{Agence d'{\'E}valuation de la Recherche et de l'Enseignement Sup{\'e}rieur}} (AERES).
In the 2008 evaluation, a method was used which is considered more precise than 
that used previously and this facilitates comparison with the British approach.
However, since only  10 traditional universities were evaluated, 
the amount of data available for the French system is lower than for the UK equivalent. 
Furthermore, only a global mark is attributed to cumulated research groupings 
so a fine-grain analysis at the level of the research groups is not possible and no transition point
is measureable. Nonetheless, we translate the AERES grades A+, A, B, C into 4*, 3*, 2*, and 1* and 
analyse the French system for hard sciences and life sciences to compare to the  British equivalent. 
In Fig.~5, the standardized success rates $\sigma = (s-\bar{s})/\sigma_s$ are
plotted against the standardized group sizes  $\nu =(N-\bar{N})/\sigma_N$ for both systems,
where $\bar{N}$ and $\bar{s}$ are the mean $N$ and $s$ values, respectively, 
and $\sigma_N$ and $\sigma_s$ are their standard deviations. A convincing degree of overlap is evident.

Palla et al. analysed international networks of co-authorship in condensed matter 
physics and found a difference in the endurance rates of teams
with above about 20 members and teams with fewer members \cite{PaBa07}. 
Endurance over time is not the same as quality, but it may serve as a rough guide
in the sense that unsuccessful research collaborations are 
unlikely to endure and vice versa.
It is therefore satisfactory to note that   our estimate  $N_c=25.4 \pm 4.8$ for physics
is consistent with the observation of Ref.~\cite{PaBa07}.

\section{Discussion}

\begin{figure}[t]
\begin{center}
\includegraphics[width=0.65\columnwidth, angle=0]{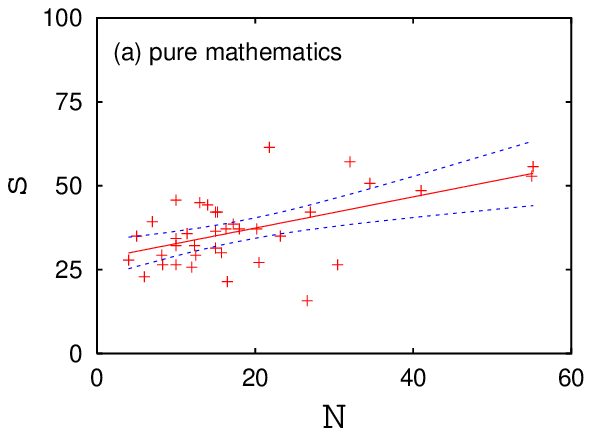}\\
\includegraphics[width=0.65\columnwidth, angle=0]{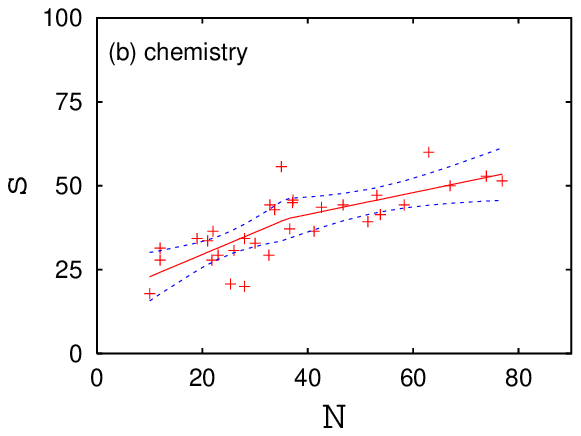}
\caption{Assessed group quality as a function of size for 
 (a) pure mathematics  and (b) chemistry. The former is best described by a single straight line.}
\end{center}
\label{chemmaths}
\end{figure}
In analysing the relationship between quality in an activity and quantity of parcipants,
we have considered communities as consisting of individuals collected as subgroups within groups.
It is a simple matter to extend these considerations to ever larger collectives, formed
from further hierarchies of communities  with self-similar 
interactions between them. 
In the case where $b_2 \ne 0$, the  quality plateau may occur at a larger value $N=N_u$, which 
marks the point where inter-subgroup interactions saturate. In this case, (\ref{Nc}) is replaced by
\begin{equation}
 s = \left\{ \begin{array}{ll}
             a_1 + b_1 N &  {\mbox{if $N \le N_c$}} \\
             a_2 + b_2 N &  {\mbox{if $N_c \le N \le N_u$}} \\
             a_3         &  {\mbox{if $N \ge N_u$}}.
             \end{array}
     \right.
\label{Nu}
\end{equation}
This double-breakpoint scenario leads to a richer phase diagram, obtainable by consideration of transfer
of $M$ nodes from group~II, of size $N_{\rm{II}}>N_u$. This is plotted in Fig.~1(c)
for a super-critical group~I. 
In Ref.~\cite{GuDa03}, where a study of the hierarchical structure of nested communities in 
a university network indicated self-organization into a self-similar structure, it was 
suggested  that some universal mechanism could be behind the 
evolution of social networks. The mechanism we propose is of this type.

\begin{figure}[t]
\begin{center}
\includegraphics[width=0.65\columnwidth, angle=0]{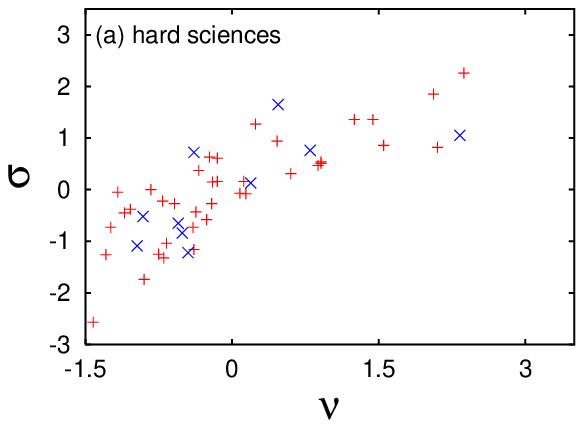}\\
\includegraphics[width=0.65\columnwidth, angle=0]{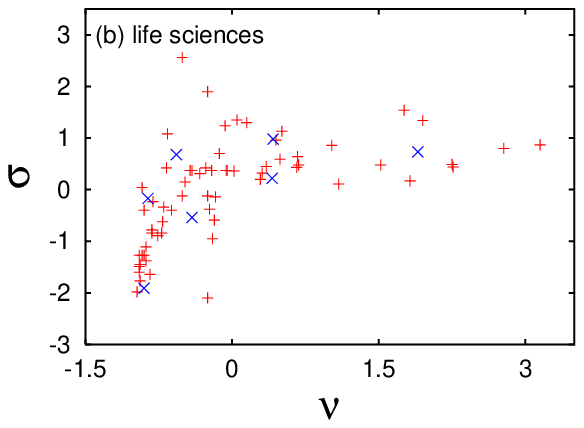}
\caption{ 
Standardized success rates $\sigma  = (s - {\bar{s}})/\sigma_s$ plotted against 
standardized sizes $\nu = (N - {\bar{N}})/\sigma_N$ for France's AERES and the UK's RAE for
(A) the hard sciences and (B) the life sciences.
The French data 
correspond to  the symbols $\times$ (blue online) and the integrated British data to $+$ (red online). }
\end{center}
\end{figure}

We have shown that there is a somewhat surprising  relationship between quality and 
quantity: in the language of statistical physics, quality is an extensive concept. 
Our study also shows that it is unwise to judge a research group solely on the basis of its quality profile - 
precisely because of this strong size dependency: small- and medium-size groups should not be expected to yield
the same quality profiles as large ones, and to compare small/medium groups to the average success rate over 
all research groups in a discipline can be misleading. 
Indeed,  Fig.~6 contains a plot of {\emph{renormalized}} research quality measures $s-\left\langle{s}\right\rangle$,
where $\left\langle{s}\right\rangle$ is the $N$-dependent expected quality value from Eq.(\ref{Nc}).
The standard deviation and range corresponding to this plot are $6.4$ and $28.8$, respectively.
These values are about half the corresponding values of $12.6$ and $52.9$, respectively, for Fig.~1(a), 
where the data are compared to their global average. Similar results are obtained for other disciplines.

Our model is based on the notion that research groups are complex systems and 
scientific quality is dominated by interactions between group members rather than by the accumulated node strengths.  
While the data analysis supports this hypothesis, one could, of course, consider
other possible mechanisms which drive the quality of group activity, such as preferential recruitment of quality to quality
(allowing for the average quality of 
individuals at some institutions to be higher than at others)
or one in which increased quality drives increased quantity (the opposite causal mechanism to that of our model).
Or there could be a bias in both the RAE and AERES where assessment of research may favour larger groups. 
However, one would expect such mechanisms to lead to a sustained increase of quality with quantity and it is therefore
hard to see how these 
alone could account for the existence and properties 
of the breakpoint, which we have clearly established through our statistical analyses.
Moreover, similar analyses of other fields confirm the existence of a breakpoint in
areas far from the natural sciences reported here, {\it{e.g.}\/}, European languages, archaelogy, philosophy and theology,
which are unlikely to have the same assessment traditions or bias as the hard or life sciences.
Of course, improvement of the model could introduce noise and bias, but very interestingly, these effects are
an order of magnitude below the basic collaborative effect.

An analysis of the type presented herein may therefore assist in the determination of which groups are, to use a boxing analogy, 
punching above or below their weight within a research arena and should be taken into account by decision makers when 
comparing research groups and when formulating strategy.
Furthermore, having established a correlation between quality and quantity, 
and ascribing this correlation as primarily due to two-way communication links, 
it is clear that facilitation of such communication should form an important management policy in academia. 
Indeed we have demonstrated that to optimize overall research quality in a given discipline,
medium-size groups should be promoted while small ones must endeavour to attain critical mass.

To summarize, inspired by mean field theory and the de Gennes polymer picture, we have developed a 
very simple model which 
relates the quality of an activity to the quantity of participants engaged in it.
In spite of its simplicity, this self-similar model captures the essential features
of the empirical data. It involves the treatment of  collectives as complex systems and 
cooperative behaviour arising from inter-nodal interactions dominate group quality.
Furthermore, these hierarchical  interactions drive ``phase transitions'' between what may be considered small, 
medium and large groups and lead to a rather rich phase structure. 
In particular, quality saturates beyond a certain group size which is related by a simple scaling relation
to a critical mass, beneath which a group is vulnerable.
Considering academia as an example of an activity where quality measurements are readily available
and where quantities of participants are known, we present saturation points and 
critical masses for research areas in a number of scientific disciplines.
Critical masses for other subjects area will be presented elsewhere.

\begin{figure}[!t]
\begin{center}
\includegraphics[width=0.65\columnwidth, angle=0]{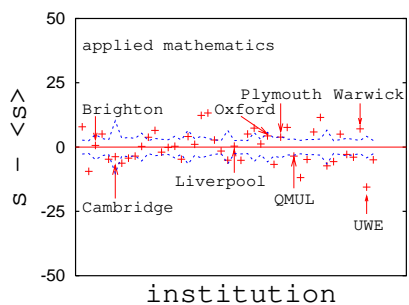}\\
\caption{ 
 The renormalized research quality $s-\left\langle{s}\right\rangle$ for applied mathematics.
The standard deviation and range corresponding to  this  plot  are about half those corresponding to Fig.1(a).}
\end{center}
\end{figure}

\acknowledgments

We thank Neville Hunt for inspiring discussions and for help with the statistical analyses. 
We also thank Arnaldo Donoso, Christian von Ferber, Housh Mashhoudy and Andrew Snowdon for comments and suggestions. 
We are grateful to Claude Lecomte, Scientific Delegate  at the 
{\emph{Agence d'{\'E}valuation de la Recherche et de l'Enseignement Sup{\'e}rieur}\/}, 
for discussions on the work of that agency.

\end{document}